\documentclass{article}
\usepackage{amsfonts,amssymb, amsmath}
\usepackage[english]{babel}

\DeclareMathAlphabet{\mathpzc}{OT1}{pzc}{m}{it}

\def\nn{\nonumber }
\def\bq{ \begin{equation} }
\def\eq{ \end{equation} }
\def\ben{ \begin{eqnarray} }
\def\en{ \end{eqnarray} }

\newtheorem{prop}{Proposition}

\newtheorem{defin}{Definition}

\newtheorem{re}{Remark}
\newenvironment{rem}{\begin{re} \rm }{\end{re}}

\newtheorem{exa}{Example}
\newenvironment{exam}{\begin{exa} \rm }{\end{exa}}

\begin{document}


\title{On the superintegrable Richelot systems. }
\author{ A V Tsiganov \\
\it\small
Saint-Petersburg State University, St.Petersburg, Russia\\
\it\small e--mail:  tsiganov@mph.phys.spbu.ru}

\date{}
\maketitle

\begin{abstract}
We introduce the  Richelot class of superintegrable systems in $N$-dimensions whose $n\leq N$ equations of motion coincide with the Abel equations on $n-1$ genus hyperellipic curve. The corresponding additional   integrals of motion are the second order polynomials of momenta and multiseparability of the Richelot superintegrable systems is related with classical theory of covers of the hyperelliptic curves.
\end{abstract}


\vskip0.1truecm
\begin{flushright}
In the antique rarely-read collections of scientific societies as well\\ 
as in comprehensive scientific correspondence of the scientist\\ 
of the past an enormous quantity of scientific matter is contained,\\ 
from which anyone capable can find something motivating to start\\ 
their own work, as well as simultaneously learn something useful.\\
\vskip0.3truecm
K. Weierstrass, "The speech delivered upon assuming the position \\
of Rector of Berlin University on October 15, 1873", Phys. Usp. 42 1219 (1999)
\end{flushright}
\vskip0.1truecm

\section{Introduction}
\setcounter{equation}{0}

In classical mechanics, superintegrable systems are characterized by the fact that they possess
more than $N$ integrals of motion functionally independent, globally defined in a $2N$-dimensional phase space. In particular, when the number of integrals is $2N-1$, the systems are said to be maximally superintegrable. The dynamics of these systems is particularly interesting: all
bounded orbits are closed and periodic \cite{bert73}. The phase space topology is also very
rich: it has the structure of a symplectic bifoliation, consisting of the usual
Liouville-Arnold invariant fibration by Lagrangian tori and of a (coisotropic) polar foliation \cite{neh72}.

The notion of superintegrability possesses an interesting analog in quantum mechanics. Sommerfeld and Bohr were the first to notice that systems allowing separation of variables in more than one coordinate system may admit additional integrals of motion. Superintegrable systems show accidental degeneracy of the energy levels, which can be removed by taking into account the quantum numbers associated to the additional integrals of motion, some of their bound state energy levels may be calculated algebraically and the corresponding wave functions are expressed in terms of polynomials. One of the best examples of this phenomenon is provided by the harmonic oscillator and the Kepler-Coulomb problem. A large number of papers have been published on super-integrability in these last years, most of them related with  second-order integrals of motion (see \cite{bal07,cd06,ev90,w65,ts09a,kal05,rodw08,ts08a,ts08b} for some recent results and an extensive list of references).

A systematic investigations of superintegrable systems have a very long story, which
began in 1761  when  Euler proposed  construction of the additional algebraic integral  for the  differential equation
 \[
\dfrac{\,d\mathrm x_1}{\sqrt{f(\mathrm x_1)}}\pm\dfrac{\,d\mathrm x_2}{\sqrt{f(\mathrm x_2)}}=0,
\]
where $f$ is an arbitrary quartic \cite{eul68}. The corresponding superintegrable St\"ackel systems have been classified in \cite{ts09a}.

The Abel theorem may be regarded as a generalization of these Euler results.  Remind that the Abel equations
\bq\label{a-eq}
\sum_{j=1}^n \dfrac{u_{i}(\mathrm x_j)\,d\mathrm x_j}{\sqrt{f(\mathrm x_j)}}=0,\qquad {i}=1,\ldots,p,
\eq
play a pivotal role in classical mechanics and that there are two approaches to investigation of the Abel equations associated  with the Jacobi and Richelot, respectively (see thirtieth lecture in the Jacobi book \cite{jac36}). In modern mathematics, the first approach or the Abel-Jacobi map is one of the main  constructions of algebraic geometry which relates an algebraic curve to its Jacobian variety. The second approach yields addition theorems theory,
moduli theory (modular equations), cryptography  and so on.

The aim of this note is to discuss the Richelot construction of  addition integrals for the Abel equations and construction of the corresponding $N$-dimensional  superintegrable systems in classical mechanics. We treat only classical superintegrable systems here, though the corresponding  results for the quantum systems follow easily.

The paper is organized as follows. In Section II, the
main Richelot results are briefly reviewed. Then we discuss
possible application of these results to classification
of the superintegrable St\"ackel systems. In Section III, the classification of
superintegrable systems separable in orthogonal coordinate systems is
treated and solved. Some open problems are discussed in the final Section.

\section{The Richelot  superintegrable systems}
\setcounter{equation}{0}
In this section we use the original Richelot notations \cite{rich42}.

Let $\mathrm y$ be the algebraic function of $\mathrm x$ defined by an equation of the form
\bq\label{g-curve}
\Phi(\mathrm x,\mathrm y)=\mathrm y^m+f_1(\mathrm x)\mathrm y^{m-1}+\cdots+f_m(\mathrm x)=0,
\eq
where $f_1(\mathrm x),\ldots,f_m(\mathrm x)$ are rational polynomials in $\mathrm x$.
According to the Abel theorem a system of the $p$ differential equations
\[
\dfrac{du_i}{d\mathrm x_1}\,d\mathrm x_1+\cdots+\dfrac{du_i}{d\mathrm x_N}d\mathrm x_N=0,\qquad i=1,\ldots,p
\]
have additional algebraic integrals if $N>p$ and if $u_1,\cdots,u_{p}$ being a set of linearly independent abelian integrals of the first kind on algebraic curve (\ref{g-curve}).

The problem of the determining of these integrals consists only in the expression of the fact that $\mathrm x_1,\ldots,\mathrm x_N$ constitute a set belonging to a lot of coresidual sets of places, so we have some determinant representations for additional integrals.

For the particular forms of the curve (\ref{g-curve}) there are some explicit formulae due by Euler \cite{eul68}, Lagrange \cite{lag}, Jacobi \cite{jac42}, Richelot \cite{rich42}, Weierstrass \cite{w} and some other \cite{bak97,cal,gr}.

\subsection{The Richelot integrals }
Following to  Richelot \cite{rich42} we will consider  hyperelliptic curve
\bq\label{h-curve}
\mathrm y^2=f(\mathrm x)\equiv A_{2n}\mathrm x^{2n}+A_{2n-1}\mathrm x^{2n-1}+\cdots+A_1 \mathrm x+A_0
\eq
and the following system of $n-1$ differential equations
\ben
&&\dfrac{d\mathrm x_1}{\sqrt{f(\mathrm x_1)}}+\dfrac{d\mathrm x_2}{\sqrt{f(\mathrm x_2)}}+\cdots+\dfrac{d\mathrm x_n}{\sqrt{f(\mathrm x_n)}}=0,\nn\\
&&\dfrac{\mathrm x_1d\mathrm x_1}{\sqrt{f(\mathrm x_1)}}+\dfrac{\mathrm x_2d\mathrm x_2}{\sqrt{f(\mathrm x_2)}}+\cdots+\dfrac{\mathrm x_nd\mathrm x_n}{\sqrt{f(\mathrm x_n)}}=0,\nn\\
&&\cdots\cdots\cdots\cdots\cdots\cdots\cdots\cdots\cdots\cdots\cdots\cdots\cdots\cdots\label{r-eq}\\
&&\dfrac{\mathrm x_1^{n-2}d\mathrm x_1}{\sqrt{f(\mathrm x_1)}}+\dfrac{\mathrm x_2^{n-2}d\mathrm x_2}{\sqrt{f(\mathrm x_2)}}+\cdots+\dfrac{\mathrm x_n^{n-2}d\mathrm x_n}{\sqrt{f(\mathrm x_n)}}=0\,.
\nn
\en
Let $a_k$ be the values of $\mathrm x$ at the branch points of the curve (\ref{h-curve}) and $F(\mathrm x)=(\mathrm x-\mathrm x_1)(\mathrm x-\mathrm x_2)\cdots(\mathrm x-\mathrm x_n)$, then in generic case  additional integrals of the Abel equations (\ref{r-eq})  are equal to
\bq\label{r-int1}
C_k=\dfrac{\left[\dfrac{\sqrt{f(\mathrm x_1)}}{F'(\mathrm x_1)}\cdot\dfrac{1}{a_k-\mathrm x_1}+\cdots+\dfrac{\sqrt{f(\mathrm x_n)}}{F'(\mathrm x_n)}\cdot\dfrac{1}{a_k-\mathrm x_n}\right]^2}
{\left[\dfrac{\sqrt{f(\mathrm x_1)}}{F'(\mathrm x_1)}+\cdots+\dfrac{\sqrt{f(\mathrm x_n)}}{F'(\mathrm x_n)}\right]^2-A_{2n}}\,F(a_k)
\eq
If $A_{2n}=0$ additional integrals of equations (\ref{r-eq}) look like
\bq\label{r-int2}
C_k=\left[\dfrac{\sqrt{f(\mathrm x_1)}}{F'(\mathrm x_1)}\cdot\dfrac{1}{a_k-\mathrm x_1}
+\cdots+\dfrac{\sqrt{f(\mathrm x_n)}}{F'(\mathrm x_n)}\cdot\dfrac{1}{a_k-\mathrm x_n}\right]^2
\,\sqrt{F(a_k)}\,.
\eq
There are $n-1$ functionally independent integrals of motion $C_k$ and, of course, their combinations are integrals of motion too.

Using special combinations of $C_k$ we can avoid calculations of
the values $a_k$ of $x$ at the branch points \cite{jac42,rich42,w}. As an example,
in his paper Richelot found  the following two algebraic integrals
\bq\label{r-pint1}
K_1=\left[\dfrac{\sqrt{f(\mathrm x_1)}}{F'(\mathrm x_1)}
+\cdots+\dfrac{\sqrt{f(\mathrm x_n)}}{F'(\mathrm x_n)}\right]^2-A_{2n-1}(\mathrm x_1+\cdots+\mathrm x_n)-A_{2n}(\mathrm x_1+\cdots+\mathrm x_n)^2
\eq
and
\bq\label{r-pint2}
K_2=\left[\dfrac{\sqrt{f(\mathrm x_1)}}{\mathrm x_1^2F'(\mathrm x_1)}
+\cdots+\dfrac{\sqrt{f(\mathrm x_n)}}{\mathrm x_n^2F'(\mathrm x_n)}\right]^2\mathrm x_1^2\mathrm x_2^2\cdots \mathrm x_n^2
-A_1\left(\dfrac{1}{\mathrm x_1}+\cdots+\dfrac{1}{\mathrm x_n}\right)-A_0\left(\dfrac{1}{\mathrm x_1}+\cdots+\dfrac{1}{\mathrm x_n}\right)^2\,.
\eq
The generating function of additional integrals was proposed by Weierstrass \cite{w},
see \cite{bak97} for detail.

\subsection{Construction of the Richelot  superintegrable systems}
Let as apply the Richelot construction  to
classification of the superintegrable systems in classical mechanics.

\begin{defin}
The $N$-dimensional integrable system  is the superintegrable Richelot system
if $n-1$, $1<n\leq N$, equations of motion   are the Abel-Richelot equations (\ref{r-eq}).
\end{defin}
It's easy to get a lot of such superintegrable Richelot systems in framework of the Jacobi separation of variables method, see \cite{ts09a,ts08a,ts08b}.

Let us start with the maximally superintegrable Richelot systems at $N=n$.
In this case construction  consists of the
one hyperelliptic curve (\ref{h-curve})
\bq\label{hc2}
\mu^2=f(\lambda), \qquad\mbox{where}\qquad f(\lambda)=A_{2n}\lambda^{2n}+A_{2n-1}\lambda_i^{2n-1}+\cdots+A_1\lambda+A_0,
\eq
and $n$ arbitrary substitutions
\bq\label{subs-lm}
\lambda_j=v_j(q_j) \qquad \mu_j=u_j(q_j)p_j,\qquad j=1,\ldots,n,
\eq
where $p$ and $q$ are canonical variables $\{p_j,q_i\}=\delta_{ij}$.

The $n$ copies of this hyperelliptic curve and these substitutions give us $n$ separated relations
\bq\label{sep-rel-l}
p_j^2\, u_j^2(q_j)=A_{2n}v_j(q_j)^{2n}+A_{2n-1}v_j(q_j)_i^{2n-1}+\cdots+A_1v_j(q_j)+A_0,\qquad j=1,\ldots,n,
\eq
where $2n+1$ coefficients $A_{2n},\ldots,A_0$ are linear functions of $n$  integrals of motion $H_{1},\ldots,H_n$ and $2n+1$  parameters $\alpha_0,\ldots,\alpha_{2n+1}$.

Solving these separated equations with respect to $H_k$ one gets
functionally independent integrals of motion
\bq
H_k=\sum_{j=1}^n ( S^{-1})_{jk}\Bigl(p_j^2+U_j(q_j)\Bigr)\,,\qquad k=1,\ldots,n=N,
\label{fint}
\eq
where $U_j(q_j)$ are so-called St\"ackel potentials and $S$ is the St\"{a}ckel matrix \cite{st95}.

If  $H_1$ is the Hamilton function, then  coordinates
$q_j(t,\alpha_1,\ldots,\alpha_n)$ are determined from the Jacobi equations
\bq
\sum_{j=1}^n\int \dfrac{S_{1j}(q_j) dq_j}
{\sqrt{\sum_{k=1}^n \alpha_k S_{1j}(q_j)-U_j(q_j)}}=\tau-t\,,
\eq
 and
\bq
\sum_{j=1}^n\int\dfrac{S_{ij}(q_j) dq_j}
{\sqrt{\sum_{k=1}^n \alpha_k S_{kj}(q_j)-U_j(q_j)}}=\beta_i\,,
\qquad i=2,\ldots,n\,,\label{stinv}
\eq
where $t$ is the time variable conjugated to the Hamilton function $H_1$. According  to
Jacobi \cite{jac36}  these equations are another form of the Abel equations (\ref{a-eq})  and
describe inversion of the corresponding Abel map.

In order to use the Richelot results   we have to impose some
constraints  on the entries of the St\"ackel matrix $S_{kj}(q_j)$, which give rise to some restrictions on the  coefficients $A_k$  \cite{ts09a,ts08a}.

Namely,  if we compare $n-1$ equations (\ref{r-eq}) and equations (\ref{stinv})
at $\lambda=\mathrm x$ one gets that the St\"akel matrix in $\lambda$ variables
has to be one of the following matrices
\bq\label{st-mat}
S^{(k)}=\left(
  \begin{array}{cccc}
    \lambda_1^k & \lambda_2^k & \cdots & \lambda_n^k \\
    \lambda_1^{n-1} & \lambda_2^{n-1} & \cdots & \lambda_n^{n-1} \\
    \lambda_1^{n-2} & \lambda_2^{n-2} & \cdots & \lambda_n^{n-2} \\
    \vdots & \vdots & \ddots & \vdots \\
    1 & 1 & \cdots & 1
  \end{array}
\right),\qquad k=n,n+1,\ldots,2n,
\eq
so that
\bq\label{r-curve2}
\mu^2=f(\lambda)=\lambda^kH_1+\lambda^{n-1}H_{n-1}\cdots+H_{n-1}\lambda+H_n+\sum_{j=0}^{2n} \alpha_j\lambda^{j}\,.
\eq
Such as $k$ is arbitrary number from $n$ to $2n$ we have a family of the  dual St\"ackel systems associated with one hyperelliptic curve (\ref{hc2}) and different blocks
of the corresponding Brill-Noether matrix  \cite{ts99,ts99a}

\begin{rem}
For any two dual systems with Hamiltonians $H_1$ and $\widetilde{H}_1$ the corresponding St\"ackel matrices $S^{(k)}$ and ${S}^{(\widetilde k)}$ are distinguished on the first row only.  These St\"ackel systems are related by  canonical transformation of the time $t\to\widetilde{t}$:
\bq\label{t-change}
\widetilde{H}_1=\mathrm v(q)\,{H_1},\quad d\widetilde{t}=\mathrm v(q)\,dt,\qquad\mbox{where}\qquad \mathrm v(q)=\dfrac{\det{S^{(k)}}}{\det{S^{(\widetilde k)}}}.
\eq
Such dual systems have common trajectories with different parametrization by the time \cite{ts99a,kal05}. Existence of the such dual systems is related with the fact that the Abel map is surjective and generically injective.
\end{rem}

\begin{rem}
For the dual systems the corresponding hyperelliptic curves (\ref{r-curve2})
are related by permutation of one of the  $\alpha$'s and Hamiltonian $H_1$ and, therefore,  such transformations are called the coupling constant metamorphoses \cite{bkm86,hi84,ts99a}. Such transformations are related with the reciprocal transformations as well \cite{ab}.
\end{rem}

\begin{rem}
There exist the Richelot superintegrable systems  that can
be solved via separation of variables in more than one coordinate system.
These systems are  associated with non-isomorphic curves  whose Jacobians are isomorphic to one another (either Jacobian of (\ref{h-curve}) could be isomorphic to a strata of another Jacobian or Jacobian of (\ref{h-curve}) could be  isogenous to a product of some different curves etc).  Such curves  have already occurred in the work of Hermite, Goursat, Burkhardt, Brioschi, and Bolza, see
Krazer \cite{kr} and a lot of modern works on  the Frey-Kani covers.

\end{rem}

Now let as briefly  consider construction of the superintegrable Richelot systems for which
$n-1$ equations of motion among the $N$ equations of motion are the Abel-Richelot equations only.
In this case to $n$ separated relations (\ref{sep-rel-l})  have to be complimented by $N-n$
separated relations
\[
\Phi_m(p_m,q_m,H_1,\ldots,H_N)=0,\qquad n<m\leq N.
\]
Solving this complete set of the separated equations with respect to integrals of motion $H_k$
we have to get $N$ functionally independent integrals of motion (\ref{fint}). As above the
Abel equations have to coincide with the Richelot equations (\ref{r-eq})  and, therefore,
the $n\times n$ block of the $N\times N$ St\"ackel matrix has to be matrix as (\ref{st-mat}).
If we take into account all these restrictions one gets complete classifications of
the superintegrable St\"ackel-Richelot systems.

The main problem is that we want to get  Hamiltonians $H_j$
in some physical variables $x$ instead of Hamiltonians (\ref{fint}) in terms of
the abstract separated variables $q$. According to \cite{ts09a,ts08a,ts08b} it leads to some additional restrictions on the coefficients $A_j$ in (\ref{hc2})
and substitutions (\ref{subs-lm}).

It easy to see that the St\"ackel integrals of motion $H_k$ (\ref{fint})
and the  Richelot additional integrals of motion are the second order polynomials in momenta
\bq\label{r-pint11}
K_1=\left[\dfrac{u_1p_1}{F'(v_1)}
+\cdots+\dfrac{u_np_n}{F'(v_n)}\right]^2-A_{2n-1}(v_1+\cdots+v_n)-A_{2n}(v_1+\cdots+v_n)^2
\eq
and
\bq\label{r-pint21}
K_2=\left[\dfrac{u_1p_1}{v_1^2F'(v_1)}
+\cdots+\dfrac{u_np_n}{v_n^2F'(v_n)}\right]^2v_1^2v_2^2\cdots v_n^2
-A_{1}\left(\dfrac{1}{v_1}+\cdots+\dfrac{1}{v_n}\right)-A_0\left(\dfrac{1}{ v_1}+\cdots+\dfrac{1}{v_n}\right)^2\,.
\eq
Here $u_j$ and $v_j$ are functions on coordinates only.

So, in the St\"ackel-Richelot case all the integrals of motion are the second order polynomials
in momenta and it allows us to find natural Hamiltonian superintegrable systems on the
Riemannian manifolds  using well-studied theory of the orthogonal coordinate systems and the corresponding Killing tensors \cite{ben,ei34,kal,stef02}.

\section{The Richelot systems separable in orthogonal coordinate systems}
All the orthogonal separable coordinate systems can be viewed as an orthogonal
sum of certain basic coordinate systems \cite{ben,ei34,kal,stef02}. Below we consider some of these
basic coordinate systems in the $n$-dimensional Euclidean space only.
\subsection{The basic orthogonal coordinate systems}
\begin{defin}
The elliptic coordinate system $\{q_i\}$ in the $N$-dimensional Euclidean space $\mathbb E_N$
with parameters $e_1 <e_2 < \cdots < e_N$
is defined through the equation
\bq\label{ell-c}
e(\lambda)=1+\sum_{k=1}^N\dfrac{x_k^2}{\lambda-e_k}=
\dfrac{\prod_{j=1}^N(\lambda-q_j)}{\prod_{i=1}^N(\lambda-e_i)}\,.
\eq
\end{defin}
The defining equation (\ref{ell-c}) should be interpreted as an identity with respect
to $\lambda$.

It is possible to degenerate the elliptic  coordinate systems in a proper way by
letting two or more of the parameters $e_i$ coincide. Then the ellipsoid will become
a spheroid, or even a sphere if all parameters coincide. Rotational symmetry
of dimension $m$ is thus introduced if $m + 1$ parameters coincide.

\begin{exam}
As an example when $e_1$ = $e_2$, we have
\bq\label{d-eq1}
e(\lambda)=1 + \dfrac{r^2}{\lambda-e_1}+\sum_{i=3}^N\dfrac{x_i^2}{\lambda-e_i}=
\dfrac{\prod_{i=1}^{N-1}(\lambda-q_i)}{\prod_{j=1}^{N-1}(\lambda-e_j)},\qquad r^2=x_1^2+x_2^2\,.
\eq
It defines elliptic coordinate system in $\mathbb E_{N-1}=\{r,x_3,\cdots,x_N\}$.
In order to get an orthogonal coordinate system $\{q_1,\cdots,q_N\}$ in $\mathbb E_N$, we could complement $r$ with an
angular coordinate $q_N$ in the $\{x_1, x_2\}$-plane, for instance through
\bq\label{d-eq2}
x_1 = r\cos q_N,\qquad x_2 = r \sin q_N\,,\qquad \mbox{where}\quad
r=\sqrt{\left.\mbox{res}\right|_{\lambda=e_1} e(\lambda)}\,.\eq
At $N=3$ these equations define the prolate spherical coordinate system.
\end{exam}
When $e_1= e_2=\cdots = e_n$ the only remaining coordinate is $r =\sqrt{\sum x_i^2\,}$
and $N-1$ angular coordinates have to be introduced on the unit sphere $\mathbb S_{N-1}$.
According to  \cite{kal} these angular coordinates are so-called ignorable coordinates.

\begin{defin}
The parabolic coordinate system $\{q_i\}$ in  $\mathbb E_N$
with parameters $e_1 <e_2 < \cdots < e_{N-1}$
is defined through the equation
\bq
e(\lambda)=\lambda-2x_N-\sum_{k=1}^{N-1}\dfrac{x_k^2}{\lambda-e_k}=
\dfrac{\prod_{j=1}^N(\lambda-q_j)}{\prod_{i=1}^{N-1}(\lambda-e_i)}.
\eq
\end{defin}
This orthogonal coordinate system can, in fact, be derived from the elliptic
coordinate system as well. Namely, substitute
\[x_i = \dfrac{x'_i}{\sqrt{e_i}},\qquad i = 1,\ldots, N-1,\qquad x_N = \dfrac{x'_N-e_N}{\sqrt{e_N}}
\]
into the (\ref{ell-c}) and let $e_N$ tend to infinity, then drop the primes one gets the parabolic coordinate system.

The parabolic coordinate system can be degenerated in the same way as the
elliptic coordinate system.

\begin{exam}
If $e_1$ = $e_2$, we have
\bq\label{d-eq3}
e(\lambda)=\lambda-2x_N-\dfrac{r^2}{\lambda-e_1}-\sum_{k=3}^{N-1}\dfrac{x_k^2}{\lambda-e_k}=
\dfrac{\prod_{j=1}^{N-1}(\lambda-q_j)}{\prod_{i=1}^{N-2}(\lambda-e_i)},\qquad r^2=x_1^2+x_2^2\,.
\eq
As above in order to get an orthogonal coordinate system $\{q_1,\cdots,q_n\}$
in $\mathbb E_N$, we could complement $r$ with an angular or ignorable  coordinate $q_N$ in the $\{x_1, x_2\}$-plane defined by (\ref{d-eq2}). At $N=3$ it is so-called rotational parabolic coordinates.
\end{exam}

\begin{defin}
The elliptic coordinate system $\{q_i\}$ on the sphere $\mathbb S_{N}$
with parameters $e_1 <e_2 < \cdots < e_{N+1}$
is defined through the equation
\bq\label{ell-sph}
e(\lambda)=\sum_{k=1}^{N+1}\dfrac{x_k^2}{\lambda-e_k}=
\dfrac{\prod_{j=1}^{N}(\lambda-q_j)}{\prod_{i=1}^{N+1}(\lambda-e_i)}\,.
\eq
\end{defin}
Notice that (\ref{ell-sph}) implies $\sum_{i=1}^{N+1} x_i^2=1$. In the similar manner
we can define elliptic coordinate system $\{q_i\}$ on the hyperboloid
$\mathbb H_{N}$ with $x_0^2-\sum_{i=1}^N x_i^2=1$ \cite{kal}. As above these
coordinates can be degenerated by letting some, but not all, parameters $e_i$ coincide.

\begin{rem}
There are some algorithms \cite{ben,stef02} and software \cite{ts05} that for a given natural Hamilton function $H=T+V$ determine if separation coordinates exist, and in that case, show how to construct them, i.e. how to get determining function $e(\lambda)$.
\end{rem}

\subsection{The maximally superintegrable Richelot systems}

The basic orthogonal coordinate systems is defined by the function
\bq
e(\lambda)=\dfrac{\prod_{i=1}^N(\lambda-q_j)}{\prod_{j=1}^M(\lambda-e_j)}
=\dfrac{\phi(\lambda)}{u(\lambda)}\,\qquad M=N,N\pm 1,
\eq
which is the ratio of the following polynomials
\bq\label{f-phi}
\phi(\lambda)=\prod_{i=1}^N(\lambda-q_j),\qquad\mbox{and}\qquad u(\lambda)=\prod_{j=1}^M(\lambda-e_j)\,.
\eq

We can  describe the maximally superintegrable Richelot systems separable in these coordinate systems using the following Proposition.

\begin{prop}
If $n=N$ separated relations have the following form
\bq\label{ell-sepr}
p_i^2\,u(q_i)^2=\dfrac{1}{2}\,\left[u(\lambda)\cdot\left( H_1\,\lambda^k+\sum_{i=2}^N H_i\,\lambda^{n-i}\,\right)-\alpha(\lambda)\right]_{\lambda=q_i},\qquad \alpha(\lambda)=\sum_{j=0}^{2N}\, \alpha_j\,\lambda^{j},
\eq
where $\alpha(\lambda)$ is arbitrary polynomial, then equations of motion (\ref{stinv}) are the Abel-Richelot equations (\ref{r-eq}).

If $k=n$ the corresponding maximally superintegrable Hamiltonian $H_1$
\[
H_1=T+V=\left.\sum_{i=1}^N \mbox{\rm res}\right|_{\lambda=q_i} \dfrac{1}{e(\lambda)}\,\cdot p_i^2-
\left.\sum_{i=1}^N\mbox{\rm res}\right|_{\lambda=q_i}\dfrac{\alpha(\lambda)}{u^2(\lambda)e(\lambda)}
\]
has a natural form in Cartesian coordinates in $\mathbb E_n$
\bq\label{pot-rich}
H_1=T+V
=\dfrac12\sum_{i=1}^N p_{x_i}^2
+\left.\sum_{i=0}^M\mbox{\rm res}\right|_{\lambda=e_i}\dfrac{\alpha(\lambda)}{u^2(\lambda)\,e(\lambda)}.
\eq
Here we introduce additional parameter $ e_0=\infty$.

 If $k> n$ then ${H}_1^{(k>n)}=\mathrm v(x)\,H_1$, where function $\mathrm v(x)$ is defined by (\ref{t-change}).
\end{prop}
It is easy to prove, that these maximally superintegrable Richelot systems coincide with the
well-known superintegrable systems \cite{bal07,cd06,ev90,w65,kal05,rodw08}.
l
For elliptic coordinate system in $\mathbb E_N$ equation (\ref{pot-rich}) yields
the following potential
\[
V=\alpha_{2N}(x_1^2+\cdots x_n^2)+\displaystyle\sum_{i=1}^N \dfrac{\gamma_i}{x_i^2},\qquad \gamma_i=\dfrac{\alpha(e_i)}{\prod_{j\neq i}(e_i-e_j)^2}\,.\]
For parabolic coordinate system in $\mathbb E_N$ one gets
\[V=\alpha_{2N}(x_1^2+\cdots 4x_N^2)+\gamma_Nx_N+\displaystyle\sum_{i=1}^{N-1} \dfrac{\gamma_i}{x_i^2},\qquad \gamma_N=4\alpha_{2N}\sum e_i+2\alpha_{2N-1}\,.\]
For elliptic coordinate system on the sphere $\mathbb S_N$ or on the hyperboloid $\mathbb H_N$
we obtain
\[
V=\displaystyle\sum_{i=1}^{N+1} \dfrac{\gamma_i}{x_i^2},\qquad \gamma_i=\dfrac{\alpha(e_i)}{\prod_{j\neq i}(e_i-e_j)^2}\,.\]

\begin{exam}
Let us consider parabolic coordinates $(q_1,q_2,q_3)$ defined by
\[
e(\lambda)=\lambda-2x_3-\dfrac{x_1^2}{\lambda-e_1}-\dfrac{x_2^2}{\lambda-e_2}=
\dfrac{(\lambda-q_1)(\lambda-q_2)(\lambda-q_3)}{(\lambda-e_1)(\lambda-e_2)}\,,
\]
whereas the corresponding momenta are equal to
\[
p_i =\dfrac{x_1p_{x_1}}{2(q_i-e_1)}+\dfrac{x_2p_{x_2}}{2(q_i-e_2)}+\dfrac{p_{x_3}}{2},\qquad i=1,\ldots,3.
\]
In this case the separated relations (\ref{ell-sepr}-\ref{ign-eq}) look like
\bq\label{par-sep}
p_i^2(q_i-e_1)^2(q_i-e_2)^2=\dfrac12\Bigl[
(H_1\lambda^2+H_2\lambda+H_3)(\lambda-e_1)(\lambda-e_2)-\alpha(\lambda)\Bigr]_{\lambda=q_i},\qquad i=1,\ldots,3.
\eq
Solving these equations with respect to $H_k$  one gets integral of motion and the following Hamilton function
\bq\label{par-H}
H_1=\dfrac{p_{x_1}+p_{x_2}+p_{x_3}}{2}+\alpha_6(x_1^2+x_2^2+4x_3^2)+\gamma_3x_3
+\dfrac{\gamma_1}{x_1^2}+\dfrac{\gamma_2}{x_2^2}+const\,.
\eq
It is maximally superintegrable Hamiltonian with the St\"{a}ckel integrals of motion $H_2,H_3$ and two additional  Richelot integral of motion $K_{1,2}$ (\ref{r-pint11}-\ref{r-pint21}):
\ben
K_1&=&\left(\,\dfrac{(q_1-e_1)(q_1-e_2)p_1}{(q_1-q_2)(q_1-q_3)}
+\dfrac{(q_2-e_1)(q_2-e_2)p_2}{(q_2-q_1)(q_2-q_3)}
+\dfrac{(q_3-e_1)(q_3-e_2)p_3}{(q_3-q_1)(q_3-q_2)}\,\right)^2\nn\\
&+&\dfrac{\alpha_5}{2}(q_1+q_2+q_3)+\dfrac{\alpha_6}{2}(q_1+q_2+q_3)^2\,,\nn\\
\label{par-K}\\
K_2&=&\left(
 \dfrac{(q_1-e_1)(q_1-e_2)p_1}{(q_1-q_2)(q_1-q_3)q_1^2}
+\dfrac{(q_2-e_1)(q_2-e_2)p_2}{(q_2-q_1)(q_2-q_3)q_2^2}
+\dfrac{(q_3-e_1)(q_3-e_2)p_3}{(q_3-q_1)(q_3-q_2)q_3^2}\right)^2q_1^2q_2^2q_3^2\nn\\
&+&\dfrac{H_3e_1+(H_3-H_2e_1)e_2}{2}\left(\dfrac{1}{q_1}+\dfrac{1}{q_2}+\dfrac{1}{q_3}\right)
-\dfrac{e_1e_2H_3}2\left(\dfrac{1}{q_1}+\dfrac{1}{q_2}+\dfrac{1}{q_3}\right)^2\,.\nn
\en
In physical variables $(x,p_x)$ these integrals have more complicated structure.

It is easy to prove that integrals $H_1,H_2,H_3$ and $K_1,K_2$ are functionally independent. of course, all these integrals of motion may be obtained in framework of the  Weierstrass approach  \cite{w} as well.
\end{exam}

\begin{exam} Now let us consider dual St\"ackel system and put $k=n+1$ in the St\"ackel matrix (\ref{st-mat}) from the previous Example. It means that we change one of the coefficients in the separated relations (\ref{par-sep}) and  consider the following separated relations
\[
p_i^2(q_i-e_1)^2(q_i-e_2)^2=\dfrac12\Bigl[
(\widetilde{H_1}\lambda^3+\widetilde{H}_2\lambda+\widetilde{H}_3)(\lambda-e_1)(\lambda-e_2)-\alpha(\lambda)\Bigr]_{\lambda=q_i},\qquad i=1,\ldots,3.
\]
Solving these equations  one gets superintegrable system with the Hamiltonian
\[
\widetilde{H}_1=\mathrm v(q)\,{H_1}=\dfrac{1}{2x_3+e_1+e_2}\,H_1,
\]
where $H_1$ is given by (\ref{par-H}). Of course, this canonical transformation of time  changes additional integrals of motion $K_{1,2}$ (\ref{par-H}).

\end{exam}

\subsection{The superintegrable Richelot systems }
Now let us consider degenerate  coordinate systems for which two or more of the parameters $e_j$ coincide.

In terms of the separated coordinates defining function $e(\lambda)$ remains meromorphic function with $n$ simple roots and $m=n,n\pm1$ simple poles. For the construction of the Richelot systems we need degenerations such that $1<n<N$.

In this case in order to get  superintegrable Richelot systems with $n-1$ additional integrals of motion we have to take $n$ separated relations (\ref{ell-sepr})
\bq\label{dell-sepr}
p_i^2\,u(q_i)^2=\dfrac{1}{2}\,\left[u(\lambda)\cdot\left( H_1\,\lambda^k+\sum_{i=2}^n H_i\,\lambda^{n-i}\right)-\alpha(\lambda)+\dfrac{1}{2}\sum_{j=n+1}^N \dfrac{u(\lambda)}{\mathrm
g_j(\lambda)} H_j\right]_{\lambda=q_i},
\eq
and $N-n$  separated relations for ignorable variables
\bq\label{ign-eq}
p_j^2=2\Bigl(\,U_j(q_j)-H_j\,\Bigr)\,,\qquad j=n+1,\ldots,N.
\eq
Here polynomials $\mathrm g_j(\lambda)$ depend on degree of degeneracy and definition of the ignorable  variables \cite{ben,kal}, whereas $U_j(q_j)$ are arbitrary functions on these ignorable (angular) variables $q_j$.

Solving
these equations with respect to integrals of motion $H_j$ one gets the  Hamilton function in the same form as (\ref{pot-rich}) in which, roughly speaking, trailing coefficient of the polynomial $\alpha(\lambda)$ depends on ignorable variables.
\begin{prop}
For degenerate elliptic or parabolic coordinates superintegrable potentials have the following form (\ref{pot-rich})
\bq\label{pot-rich2}
V=
\left.\sum_{i=0}^m\mbox{\rm res}\right|_{\lambda=e_i}\,
\dfrac{\alpha(\lambda)-U_i}{u^2(\lambda)\,e(\lambda)},\qquad\qquad e_0=\infty,
\eq
where $U_i=0$ for single roots $e_i$ of initial function $(\lambda-e_1)\cdots(\lambda-e_M)$ (\ref{f-phi}) after degeneration $e_k=e_j$. For degenerate roots $e_k=e_j$ potential $U_i$ are arbitrary functions on the corresponding ignorable variables.
\end{prop}
It allows us classify all the superintegrable Richelot systems  using known classification of the orthogonal coordinate systems \cite{bal07,cd06,ev90,w65,ts09a,kal05,rodw08}.

\begin{exam}
Let us consider prolate spherical coordinate system $(q_1,q_2,q_3)$ defined by
\[
e(\lambda)=1+\dfrac{x_1^2+x_2^2}{\lambda-e_1}+\dfrac{x_3^2}{\lambda-e_3}=
\dfrac{(\lambda-q_1)(\lambda-q_2)}{(\lambda-e_1)(\lambda-e_3)},\qquad
q_3=\arctan\left(\frac{x_1}{x_2}\right)\,.
\]
The corresponding momenta are
\[
p_1=\dfrac{x_1p_{x_1}+x_2p_{x_2}}{2(q_1-e_1)}+\dfrac{x_3p_{x_3}}{2(q_1-e_3)},\qquad
p_2=\dfrac{x_1p_{x_1}+x_2p_{x_2}}{2(q_2-e_1)}+\dfrac{x_3p_{x_3}}{2(q_2-e_3)},\qquad
p_3 =x_2p_{x_1} -x_1p_{x_2}\,.
\]
In this case $\mathrm g(\lambda)=(e_3-e_1)^{-1}(\lambda-e_1)$ and the separated relations (\ref{dell-sepr}-\ref{ign-eq}) look like
\[
\begin{array}{l}
p_{i}^2(q_{i}-e_1)^2(q_{i}-e_3)^2=
\dfrac12\left[(H_1\lambda+H_2)(\lambda-e_1)(\lambda-e_2)-\alpha(\lambda)
+\dfrac{(\lambda-e_3)(e_3-e_1)H_3}{2}\right]_{\lambda=q_{i}}\,,\\
\\
p_3=2\left(\,U\left(q_3\right)-H_3\,\right)\,,
\end{array}
\]
where $\alpha(\lambda)=\alpha_4\lambda^4+\alpha_3\lambda^3+\alpha_2\lambda^2+\alpha_1\lambda+\alpha_0$.

Solving these equations with respect to $H_k$  one gets integrals of motion and  the following Hamilton function
\[
H_1=\dfrac{p_{x_1}+p_{x_2}+p_{x_3}}{2}+\alpha_4(x_1^2+x_2^2+x_3^2)
+\dfrac{\gamma_1-U\left(\dfrac{x_1}{x_2}\right)}{x_1^2+x_2^2}+\dfrac{\gamma_3}{x_3^2}
-2\alpha_4(e_3+e_1)-\alpha_3\,,
\]
where
\[
\gamma_{1,3}=\dfrac{\alpha(e_{1,3})}{(e_1-e_3)^2}\,.
\]
It is superintegrable Hamiltonian with the St\"{a}ckel integrals of motion $H_2,H_3$ and additional Richelot integral of motion $K_1$ (\ref{r-pint11}), which  is equal to
\[
 K_1= \left(\dfrac{(q_1-e_1)(q_1-e_3)p_1}{q_1-q_2}
 +\dfrac{(q_2-e_1)(q_2-e_3)p_2}{q_2-q_1}\right)^2
 -\dfrac{(H_1-\alpha_3)(q_1+q_2)}{2}+\dfrac{\alpha_4(q_1+q_2)^2}2\,.
\]
In physical variables $(x,p_x)$ one gets the following expression for this integral of motion
\[
K_1=\dfrac{(x_1p_{x_1}+x_2p_{x_2}+x_3p_{x_3})^2}4
+\dfrac{e_1+e_3-x_1^2-x_2^2-x_3^2}2\Bigl(\alpha_4(e_1+e_3-x_1^2-x_2^2-x_3^2)+\alpha_3-H_1\Bigr)\,.
\]

The second Richelit integral $K_2$ (\ref{r-pint21}) looks like
\[
K_2=\left(\dfrac{(q_1-e_1)(q_1-e_3)p_1}{(q_1-q_2)q_1^2}
         +\dfrac{(q_2-e_1)(q_2-e_3)p_2}{(q_2-q_1)q_2^2}\right)^2\,q_1^2q_2^2
-A_1\left(\dfrac{1}{q_1}+\dfrac{1}{q_2}\right)-A_0\left(\dfrac{1}{q_1}+\dfrac{1}{q_2}\right)\,,
\]
where
\[
A_1=\dfrac12\left(
e_1e_3H_1-(e_1+e_3)H_2+(e_1-e_3)H_3-\alpha_1\right),\qquad
A_0=\dfrac12\left(e_1e_3H_2-e_3(e_1-e_3)H_3-\alpha_0\right)\,.
\]
Of course, substituting $H_1,\ldots,H_3$ into $K_2$ one gets that $K_1=K_2$, because in this case we have only one Abel-Richelot equation, i.e. $n-1=1$. It means that Hamiltonian $H_1$ in $\mathbb E_3$ does not maximally superintegrable.
\end{exam}

\begin{exam}
Let us consider rotational parabolic coordinates $(q_1,q_2,q_3)$ defined by
\[
e(\lambda)=\lambda-2x_3-\dfrac{x_1^2+x_2^2}{\lambda-e_1}=
\dfrac{(\lambda-q_1)(\lambda-q_2)}{\lambda-e_1},\qquad
q_3=\arctan\left(\frac{x_1}{x_2}\right)\,,
\]
whereas the corresponding momenta look like
\[
p_1 =\dfrac{x_1p_{x_1}+x_2p_{x_2}}{2(q_1-e_1)}+\dfrac{p_{x_3}}2,\qquad
p_2 =\dfrac{x_1p_{x_1}+x_2p_{x_2}}{2(q_2-e_1)}+\dfrac{p_{x_3}}2, \qquad
p_3 =x_2p_{x_1} -x_1p_{x_2}\,.
\]
In this case  $\mathrm g(\lambda)=(\lambda-e_1)$ and the separated relations (\ref{dell-sepr}-\ref{ign-eq}) are equal to
\[
\begin{array}{l}
p_{1,2}^2(q_{1,2}-e_1)^2=\dfrac{1}{2}\left[
(H_1\lambda+H_2)(\lambda-e_1)-\alpha(\lambda)+\dfrac{H_3}{2}\right]_{\lambda=q_{1,2}}\,,\\
\\
p_3^2=2(U(q_3)-H_3)\,,\end{array}
\]
where $\alpha(\lambda)=\alpha_4\lambda^4+\alpha_3\lambda^3+\alpha_2\lambda^2+\alpha_1\lambda+\alpha_0$.

Solving these equations with respect to $H_k$  one gets integrals of motion and the following Hamilton function
\[
H_1=\dfrac{p_{x_1}+p_{x_2}+p_{x_3}}{2}+\alpha_4(x_1^2+x_2^2+4x_3^2)+2(2\alpha_4e_1+\alpha_3)x_3
+\dfrac{\alpha(e_1)-U\left(\dfrac{x_1}{x_2}\right)}{x_1^2+x_2^2}
-3\alpha_4e_1^2-2\alpha_3e_1-\alpha_2\,.
\]
It is superintegrable Hamiltonian with the St\"{a}ckel integrals of motion $H_2,H_3$ and additional Richelot integral of motion $K_1$ (\ref{r-pint11}), which  is equal to
\ben
K_1&=&\left(\dfrac{(q_1-e_1)p_1}{q_1-q_2}+\dfrac{(q_2-e_1)p_2}{q_2-q_1}\right)^2
+\dfrac{\alpha_3}{2}(q_1+q_2)+\dfrac{\alpha_4}{2}(q_1+q_2)^2\nn\\
&=&\dfrac{p_{x_3}^2}{4}+2\alpha_4x_3^2+(2\alpha_4e_1+\alpha_3)x_3+\dfrac{e_1(\alpha_4e_1+\alpha_3)}{2}\,.
\en
As above $K_1=K_2$ (\ref{r-pint11}-\ref{r-pint21}) in this case.
\end{exam}

\begin{exam}
Let us consider degenerate elliptic coordinate system on the sphere $\mathbb S_3$ in $\mathbb E_4$, so that coordinates $(q_1,q_2,q_3)$ are defined by
\[
e(\lambda)=\dfrac{x_1^2+x_2^2}{\lambda-e_1}+\dfrac{x_3^2}{\lambda-e_3}+\dfrac{x_4^2}{\lambda-e_4}=
\dfrac{(\lambda-q_1)(\lambda-q_2)}{(\lambda-e_1)(\lambda-e_3)(\lambda-e_4)}\,,
\qquad q_3=\arctan\left(\frac{x_1}{x_2}\right)\,.
\]
It means that radius of the sphere is equal to $R=\sum_{i=1}^4 x_i^2=1$.

In this case  $\mathrm g(\lambda)=(e_3-e_1)^{-1}(e_1-e_4)^{-1}(\lambda-e_1)$ and pair of the separated relations have the common form
\ben
p_i^2(q_i-e_1)^2(q_i-e_3)^2(q_i-e_4)^2&=&\dfrac{1}{2}\,
\Bigl[(H_1\lambda+H_2)(\lambda-e_1)(\lambda-e_3)(\lambda-e_4)-\alpha(\lambda)\nn\\
&+&
(e_3-e_1)(e_1-e_4)(\lambda-e_3)(\lambda-e_4)H_3\Bigr]_{\lambda=q_{1,2}}\,,\label{sph-rel}
\en
where $\alpha(\lambda)$ is fourth order polynomial with arbitrary coefficients
and third separated relation is equal to
\[p_3^2=2(U(q_3)-H_3)\,.\]
Solving separated equations with respect to $H_k$  one gets integrals of motion and the following Hamilton function
\[
H_1=\dfrac{1}{2}\left(
\sum_{i=1}^4 x_i^2\cdot \sum_{i=1}^4 p_i^2-\left(\sum_{i=1}^4 x_ip_i\right)^2\right)
+\dfrac{\gamma_1+U\left(\frac{x_1}{x_2}\right)}{x_1^2+x_2^2}
+\dfrac{\gamma_3}{x_3^3}+\dfrac{\gamma_4}{x_4^2}-\dfrac{\alpha_4}{R}\,,\qquad \gamma_i=\dfrac{\alpha(e_i)}{\prod_{j\neq i}(e_i-e_j)^2}\,.
\]
It is superintegrable Hamiltonian and additional Richelot integrals of motion looks like
\ben
K_1&=&\left(\dfrac{(q_1-e_1)(q_1-e_3)(q_1-e_4)p_1}{q_1-q_2}+\dfrac{(q_2-e_1)(q_2-e_3)(q_2-e_4)p_2}{q_2-q_1}\right)^2\nn\\
&+&\dfrac{(e_1+e_3+e_4)H_1+\alpha_3-H_2}{2}\,\Bigl(q_1+q_2\Bigr)+\dfrac{\alpha_4-H_1}{2}\,\Bigl(q_1+q_2\Bigr)^2\,.
\en
In this case $n=2$ and, therefore, $K_1=K_2$ (\ref{r-pint11}-\ref{r-pint21}).

In this case change of the time (\ref{t-change}) at $k=n+1$ yields the following transformation of pair of the separated relation (\ref{dell-sepr})- (\ref{sph-rel})
\[
p_i^2\,u(q_i)^2=\dfrac{1}{2}\,\left[u(\lambda)\cdot\left( H_1\,\lambda^2+H_2\right)-\alpha(\lambda)+\dfrac{1}{2} \dfrac{u(\lambda)}{\mathrm
g_3(\lambda)} H_3\right]_{\lambda=q_i}=\left.\dfrac{H_1}{2}\,\lambda^5+\ldots\right|_{\lambda=q_i}\,.
\]
In the right hand side of this equations we obtain $2n+1$-order polynomial in $\lambda$ and, therefore, the corresponding pair of the Abel equations are no longer the Richelot equations (\ref{r-eq}). This change of the time preserves integrability, but destroys  superintegrability.
\end{exam}

\section{Conclusion}
According to \cite{ts09a,ts08a,ts08b} there are two classes of superintegrable systems for which
the angle variables are either \textit{logarithmic} or \textit{elliptic}  functions. In the both cases one gets additional single-valued integrals of motion using addition theorems, which are particular cases of the Abel theorem.

The main aim of this note is to discuss one of the oldest but almost completely forgotten in modern literature Richelot's approach to construction and to investigation of the superintegrable systems separable in  orthogonal coordinate systems. Of course, these $n$-dimensional superintegrable systems may be obtained using another known methods (see \cite{bal07,cd06,ev90,w65,kal05,rodw08} and references within). Nevertheless we think that new definition  (\ref{pot-rich}),(\ref{pot-rich2})
\[
V=
\left.\sum\,\mbox{res}\right|_{\lambda=e_i}\,
\dfrac{\alpha(\lambda)}{u^2(\lambda)\,e(\lambda)},\qquad u(\lambda)=\prod_{j=1}^M(\lambda-e_j)\,,
\]
of the superintegrable potentials through defining function $e(\lambda)$ of coordinate system and arbitrary polynomial $\alpha(\lambda)$ may be useful in applications.

It will be interesting to get quantum counterparts of the Richelot integrals of motion and to study the algebra of integrals of motion in the algebro-geometric terms. Another perspective consists in the classification of the Richelot superintegrable systems on the  Darboux spaces.

 One more important issue concerns relation of  multiseparability of the Richelot superintegrable systems with classical theory of covers of the hyperelliptic curves.


\begin{thebibliography}{10}

\bibitem{ab}
S. Abenda,
\newblock{\em Reciprocal transformations and local Hamiltonian structures of hydrodynamic type systems},
J. Phys. A: Math. Theor. v. 42, 095208, 20 p., 2009.

\bibitem{bak97}
 H.\,F. Baker,
\newblock{\em Abel's theorem and the allied theory including the theory of the theta functions},
Cambridge: University Press, 1897.

\bibitem{bal07} A. Ballesteros, F.J. Herranz,
\newblock{\em Universal integrals for superintegrable systems on
N-dimensional spaces of constant curvature}, J. Phys. A: Math. Theor., v. 40, pp.F51-F59, 2007.

\bibitem{ben}
S. Benenti,
\newblock{\em Orthogonal separable dynamical systems}, in "Differential Geometry
and Its Applications", Vol. I, Proceedings of the 5th International Conference on
Differential Geometry and Its Applications, Silesian University at Opava, August
24-28, 1992, O.Kowalski and D.Krupka Eds., 163-184 (1993).
\\
\\
S. Benenti, \newblock{\em Separability on Riemannian manifolds}, available at
http://www2.dm.unito.it/$\sim$benenti/ (2004).

\bibitem{bert73}
J. Bertrand, \newblock{\em Th\`{e}or\`{e}me relatif au mouvement d\'{}un point attir\'{e}vers un centre fixe},
Comptes Rendus. Acad. Sci. Paris, v.LXXVII, p.849-853, 1873.

\bibitem{bkm86}
C.P. Boyer, E.G. Kalnins, Miller W. Jr.,
\newblock{\em St\"ackel-equivalent integrable Hamiltonian systems}, SIAM
J. Math. Anal., v.17, pp.778-797, 1986.

\bibitem{cal}
A. Caley, \newblock{\em An Elementary Treatise on Elliptic Functions}, Constable and Company Ltd,  London, 1876.

\bibitem{cd06}
C. Daskaloyannis, K. Ypsilantis,
\newblock{\em Unified treatment and classification of superintegrable systems with integrals quadratic in momenta on a two-dimensional manifold}, J. Math. Phys., v.47, 042904,  2006.

\bibitem{ei34} L. P. Eisenhart, \newblock{\em Separable systems of St\"ackel}, Ann. of Math., v.35, pp. 284–305, 1934.

\bibitem{ev90} N.W. Evans,\newblock{\em Superintegrability in classical mechanics}, Phys.
Rev. A v.41, pp.5666-5676, 1990.

\bibitem{eul68}
L. Euler, \newblock{\em Institutiones Calculi integralis}, Ac.Sc. Petropoli, 1761,
(Russian translation GITL, Moskow, 1956.)

\bibitem{jac36}
C. G. J. Jacobi,
\newblock{\em Vorlesungen \"uber Dynamik}, K\"onigsberg, 1866.

\bibitem{jac42}
C. G. J. Jacobi,
\newblock{\em \"{U}ber eine neue Methode zur Integration der hyperelliptischen Differentialgleichungen und \"{u}ber die rationale Form ihrer vollst\"{a}ndigen algebraischen Integralgleichungen},
J. Reine Angew. Math. v.32, p.220-227, 1846.

\bibitem{w65}
J. Fris, V. Mandrosov, Ya.A. Smorodinsky, M. Uhlir and P.
Winternitz, \newblock{\em On higher symmetries in quantum mechanics}, Phys. Lett.
v.16, pp.354-356, 1965.

\bibitem{gr}
A.G. Greenhill,
 \newblock{\em  The applications of elliptic functions}, Macmillan and Co, London,  1892.

\bibitem{ts05}
Yu.A. Grigoryev, A.V. Tsiganov,
\newblock{\em Symbolic software for separation of variables in the Hamilton-Jacobi equation for the L-systems}, Regular and Chaotic Dynamics, v.10(4), p.413-422, 2005.

\bibitem{ts09a}
Yu. A. Grigoryev, V. A. Khudobakhshov, A.V. Tsiganov,
\newblock{\em On the Euler superintegrable systems},
J. Phys. A: Math. Theor., v.42, 075202, (11pp), 2009.

\bibitem{hi84}
 J. Hietarinta, B. Grammaticos, B. Dorizzi, A. Ramani A.,
{\em Coupling-constant metamorphosis and duality between integrable Hamiltonian systems},
Phys. Rev. Lett., v.53, pp.1707-1710, 1984.

\bibitem{kal}
E.G. Kalnins,
\newblock{\em Separation of variables for Riemannian spaces of constant curvature},
Pitman Monographs and Surveys in Pure and Applied Mathematics, 28.
Longman Scientific \& Technical, Harlow; John Wiley \& Sons, Inc., New York, 1986.
\\
\\
E. G. Kalnins and W. Miller, Jr., \newblock{\em Separation of variables on n-
dimensional Riemannian manifolds. I. The n-sphere $S_n$ and Euclidean n-
space $R_n$}, J. Math. Phys., v.27, pp. 1721–1736, 1986.

\bibitem{kal05}
E.G. Kalnins,J.M.Kress and W.Miller Jr, \newblock{\em Second-order
superintegrable systems in conformally flat spaces. I,II,III},
J.Math.Phys, v.46, 053509, 28 pages, 053510, 15 pages, 103507, 28
pages, 2005.

\bibitem{kr}
 A. Krazer,
 \newblock{\em Lehrbuch der Thetafunctionen},
 Leipzig, 1903; Chelsea Reprint, New York, 1970.


\bibitem{lag}
J.L. Lagrange,
\newblock{\em Th\'{e}orie des fonctions analytiques}, chapter 2, 1797.

\bibitem{neh72}
N. N. Nekhoroshev, \newblock{\em Action-angle variables and their generalization}, Trans. Moscow Math. Soc., v.26, p.180-198, 1972.

\bibitem{stef02}
S. Rauch-Wojciechowski, C. Waksj\"{o}, \newblock{
\em How to find separation coordinates
for the Hamilton–Jacobi equation: a criterion of
separability for natural Hamiltonian systems}, Math. Phys. Anal. Geom., v.6, pp. 301-348, 2003.

\bibitem{rich42}
F. Richelot,
\newblock{\em \"Uber die Integration eines merkw\"urdigen Systems von Differentialgleichungen},
J. Reine Angew. Math. v.23, p.354-369, 1842.

\bibitem{rodw08}
M.A. Rodr\'{i}guez, P. Tempesta, P. Winternitz,
\newblock{\em Reduction of superintegrable systems: the anisotropic harmonic oscillator},
Phys. Rev. E 78, 046608, 6 pages, 2008.


\bibitem{st95}
P. St\"{a}ckel  {\em \"Uber die Integration der
Hamilton--Jacobischen Differential Gleichung Mittelst Separation der
Variabel}, Habilitationsschrift, Halle, 1891.

\bibitem{ts99}	A.V. Tsiganov,
\newblock{\em The St\"{a}ckel systems and algebraic curves},
J. Math. Phys., v.40, p.279-298, 1999.

\bibitem{ts99a}
A.V. Tsiganov,
\newblock{\em Duality between integrable St\"{a}ckel systems},
J. Phys.A: Math. Gen., v.32, p.7965--7982, 1999.

\bibitem{ts08a}
A.V. Tsiganov,
\newblock{\em Addition theorem and the Drach superintegrable systems},
J. Phys. A: Math. Theor., v.41(33), 335204 (16pp), 2008.

\bibitem{ts08b}
A.V. Tsiganov,
\newblock{\em
Leonard Euler: addition theorems and superintegrable systems},
Regular and Chaotic Dynamics, v.14(3), pp. 389-406, 2009.









\bibitem{w}
K. Weierstrass,
\newblock{\em Bemekungen \"{u}ber die integration der hyperelliptischen differential-gleichungen}, \newblock{ Math. Werke}, v.I, p.267, Berlin, Mayer and M\"{u}ller,
 1895.

\end{thebibliography}
\end{document}